\documentclass{PoS}

\usepackage{color}

\usepackage[normalem]{ulem}  

\newcommand{\be}{\begin{equation}}
\newcommand{\ee}{\end{equation}}
\newcommand{\bea}{\begin{eqnarray}}
\newcommand{\eea}{\end{eqnarray}}

\title{Multi-winding flux tubes in CFL quark matter}

\ShortTitle{Multi-winding flux tubes in CFL quark matter}

\author{Alexander Haber\\
        Department of Physics, Washington University, St Louis, MO 63130, USA\\
        E-mail: \email{ahaber@physics.wustl.edu}}

\author{\speaker{Andreas Schmitt}\\
        Mathematical Sciences and STAG Research Centre, University of Southampton, Southampton SO17 1BJ, United Kingdom\\
        E-mail: \email{a.schmitt@soton.ac.uk}}

\abstract{Color-flavor locked quark matter can be described as a three-component superconductor 
and thus shows unconventional behavior in the transition regime from type-I to type-II superconductivity. We discuss this behavior by studying magnetic line defects in a Ginzburg-Landau approach, taking into account all possible values of the three winding numbers.
After a brief discussion of the defects that include baryon circulation we focus on pure magnetic flux tubes. We show that at strong coupling, relevant for neutron stars, type-II behavior is conceivable and the most preferred configuration has minimal total winding. Only at weak coupling we find a regime where multi-winding flux tubes are preferred, although this regime most likely requires an unrealistically large superconducting gap. }

\FullConference{XIII Quark Confinement and the Hadron Spectrum - Confinement2018\\
		31 July - 6 August 2018\\
		Maynooth University, Ireland}

\begin{document}

\section{Introduction and main results}

Sufficiently cold and dense matter is a color superconductor in the color-flavor locked (CFL) phase 
\cite{Alford:1998mk,Alford:2007xm}. CFL breaks baryon number conservation spontaneously and thus can be expected to behave as a baryon superfluid. As a consequence, rotating CFL is expected to form topologically stable quantized vortices \cite{Forbes:2001gj,Eto:2013hoa}. The energetically preferred "semi-superfluid" vortices \cite{Balachandran:2005ev,Alford:2016dco} carry magnetic flux, which is of interest for the physics of neutron stars in whose cores CFL may exist. CFL also allows for line defects which carry magnetic flux but no baryon circulation
\cite{Iida:2002ev,Iida:2004if}. These "pure" magnetic flux tubes are not protected by topology, but can be stabilized through an external magnetic field \cite{Haber:2017oqb}, and possibly may form in the evolution of neutron stars. 

All these defects can be studied within a Ginzburg-Landau approach that contains three scalar fields which are charged under one gluon field and one combination of a gluon and the photon. This 
multi-component nature puts CFL in the wider context of other unconventional superconducting systems such as 
two-band superconductors \cite{Carlstrom:2010wn} or a mixture of superfluid neutrons and superconducting protons in dense nuclear matter \cite{Buckley:2004ca,Alford:2007np,Haber:2016ljn,Haber:2017kth}. For these systems 
it was shown that the coupling of the superconductor to a second component (directly and/or through 
the gauge field if both components are charged) induces unconventional behavior regarding 
the type-II regime, where magnetic flux tubes are formed. This includes a modification of the
critical value of the Ginzburg-Landau parameter for the transition from type-I to type-II behavior, a first order transition from the Meissner phase to the phase with flux tubes, the 
existence of flux tube clusters ("type-1.5 superconductivity"), and the potential existence of multi-winding flux tubes. Here we investigate this unconventional behavior in the CFL phase. 

In some parts of these proceedings, Sec.\ \ref{sec:GL}, \ref{sec:eom}, and \ref{sec:type}, 
we recapitulate the results of Ref.\ \cite{Haber:2017oqb} 
in a compact way. In addition, we present several novel results. Firstly, in Sec.\ \ref{sec:asym}
we discuss the asymptotic behavior of line defects in CFL in a more general way than previously
done, in the presence of three potentially different components and allowing for both baryon circulation and magnetic flux. This slightly generalizes earlier studies which only considered two components \cite{Eto:2009kg} or only considered configurations without baryon circulation \cite{Haber:2017oqb}. Secondly, and more importantly, in Sec.\ \ref{sec:multi} we present a systematic 
study of multi-winding flux tubes without baryon circulation. The results of this study are as follows. We demonstrate that the new magnetic defect pointed out in Ref.\ \cite{Haber:2017oqb} is indeed the 
most preferred configuration in neutron star conditions by showing explicitly that all other configurations have a larger free energy 
per magnetic flux. Unconventional behavior regarding higher winding numbers can only be seen at small values of the strong coupling constant and under the assumption that CFL is a type-II superconductor. While this is plausible at strong coupling, at weak coupling this assumption requires a superconducting gap larger than predicted from 
perturbative QCD. Possibly there is some intermediate regime where perturbative methods fail such that type-II behavior is possible, while at the same time allowing for multi-winding flux tubes. 

In future studies it would be interesting to understand the behavior of CFL in the presence of 
an external magnetic field {\it and} rotation. It is currently unknown which of the suggested defects form in this case,
possibly a combination of them, and whether and how they arrange themselves in a regular lattice. This 
question is of high relevance for neutron stars, for instance for "color-magnetic mountains", which are ellipticities of the star sustained by a color-superconducting flux tube lattice and which are potentially large enough to lead to detectable emission of gravitational waves \cite{Glampedakis:2012qp}.

\section{Ginzburg-Landau potential}
\label{sec:GL}

The usual Ginzburg-Landau potential in the theory of superconductivity for a complex scalar field $\phi$ and 
the spatial components of a $U(1)$ gauge field ${\bf A}$ reads 
\bea \label{U0}
U &=& \left|\left(\nabla+ie{\bf A}\right)\phi\right|^2-\mu^2|\phi|^2+\lambda |\phi|^4 + \frac{{\bf B}^2}{2}\, ,
\eea
where the magnetic field is ${\bf B}=\nabla\times {\bf A}$ and $e=\sqrt{4\pi\alpha}$ is the elementary charge 
with the fine structure constant $\alpha$ (using Heaviside-Lorentz units), where $\lambda>0$ is a coupling constant, and $\mu$ can be viewed as a chemical potential. This potential can be generalized to 
more than one complex field and more than one gauge field. In dense three-color, three-flavor quark matter, the order parameter 
is a complex $3\times 3$ matrix $\Phi$ in the anti-triplet representations in color and flavor space. 
Restricting ourselves to diagonal order parameters, we have three complex scalar fields, $\Phi={\rm diag}(\phi_1,\phi_2,\phi_3)$. The homogeneous CFL state is given by 
$\phi_1=\phi_2=\phi_3$. Since we take into account electromagnetism and QCD, there are in general 9 gauge fields: 8 gluons and the photon. Having reduced the order parameter to a diagonal form, the number of gauge fields needed to discuss magnetic flux tubes in CFL is reduced to 3, one gluon field $A_\mu^3$ and two
mixtures of the eighth gluon and the photon, 
\bea
\tilde{A}_\mu^8 &=& \cos\theta \,A_\mu^8+\sin\theta\, A_\mu \, , \\[2ex]
\tilde{A}_\mu &=& -\sin\theta \,A_\mu^8+\cos\theta\, A_\mu \, , 
\eea
where the  mixing angle is given by $e$ and the strong coupling constant $g$ \cite{Schmitt:2003aa},
\be \label{costheta}
\cos\theta = \frac{\sqrt{3}g}{\sqrt{3g^2+4e^2}} \, ,  \qquad \sin\theta = - \frac{2e}{\sqrt{3g^2+4e^2}} \, .
\ee
The corresponding 
Ginzburg-Landau potential up to fourth order in the fields can then be derived from the  potential 
for $\Phi$ and generalizes (\ref{U0}) to 
\bea \label{U123}
U &=&  \left|\left(\nabla+i\frac{g}{2}{\bf A}_3+i\tilde{g}_8\tilde{\bf A}_8\right)\phi_1\right|^2
+\left|\left(\nabla-i\frac{g}{2}{\bf A}_3+i\tilde{g}_8\tilde{\bf A}_8\right)\phi_2\right|^2+\left|\left(\nabla-2i\tilde{g}_8\tilde{\bf A}_8\right)\phi_3\right|^2 \nonumber\\[2ex]
&&-\mu^2(|\phi_1|^2+|\phi_2|^2+|\phi_3|^2)+\lambda(|\phi_1|^4+|\phi_2|^4+|\phi_3|^4)-2h(|\phi_1|^2|\phi_2|^2+|\phi_1|^2|\phi_3|^2+|\phi_2|^2|\phi_3|^2) \nonumber \\[2ex]
&&+\frac{\tilde{\bf B}^2}{2} + \frac{{\bf B}_3^2}{2}+ \frac{\tilde{\bf B}_8^2}{2}\, . 
\eea
where we have neglected quark masses, and where the coupling to the rotated gluon is given by 
\be
\tilde{g}_8 = \frac{\sqrt{3g^2+4e^2}}{6}\, .
\ee
The potential (\ref{U123}) shows that the rotated magnetic 
field $\tilde{\bf B}$ plays a trivial role; it simple penetrates CFL since all scalar fields are neutral with respect to it. Therefore, the numerical calculation of magnetic flux tubes in CFL involves three scalar fields and two gauge fields. 
The parameters $\mu$, $\lambda$, $h$ can be computed within perturbative QCD, in which case they are
\cite{Giannakis:2001wz}
\bea \label{weak}
\mu^2 = \frac{48\pi^2}{7\zeta(3)}T_c(T_c-T) \, , \qquad 
\lambda =\frac{72\pi^4}{7\zeta(3)}\frac{T_c^2}{\mu_q^2}  \, ,\qquad 
h= -\frac{36\pi^4}{7\zeta(3)}\frac{T_c^2}{\mu_q^2} \, ,
\eea
where $\mu_q$ is the quark chemical potential, $T$ the temperature, $T_c$ the critical temperature of CFL, and $\zeta$ the Riemann zeta function. For the following derivations, we will keep $\mu$, $\lambda$, $h$ general. For later convenience we define 
\be
\eta \equiv \frac{h}{\lambda} \, . 
\ee
Later, in our numerical results we shall use the weak-coupling result $\eta=-1/2$. In general, boundedness of the potential requires $\eta<+1/2$, and it can be shown that, in the absence of a magnetic field,  CFL is the ground state for 
$-1<\eta<1/2$, which includes the weak-coupling value. 

\section{Equations of motion for flux tube profiles}
\label{sec:eom}

In our discussion of line defects we restrict ourselves to straight lines and thus 
work in cylidrical coordinates $(r,\varphi,z)$. We write the scalar fields in terms of their modulus and phase, 
\be \label{phi}
\phi_i({\bf r}) = \frac{\rho_i(r)}{\sqrt{2}} e^{i\psi_i(\varphi)} \, , \qquad \psi_i(\varphi) = n_i \varphi \, , 
\ee
where $n_1, n_2, n_3\in \mathbb{Z}$ are the winding numbers, which determine the baryon circulation $\Gamma$ and the magnetic fluxes $\Phi_3$ and $\tilde{\Phi}_8$, 
\be
\Gamma = \frac{\pi}{3\mu_q}\frac{n_1+n_2+n_3}{3} \, , \qquad 
\Phi_3 = \frac{2\pi}{g}(n_2-n_1)  \, , \qquad 
\tilde{\Phi}_8 = \frac{\pi}{\tilde{g}_8}\frac{2n_3-n_1-n_2}{3}  \, . 
\label{Phi8}
\ee
Nonzero baryon circulation makes the line defect a superfluid vortex, nonzero magnetic fluxes make it 
a (color-)magnetic flux tube. In general, CFL line defects have both. 

The gauge fields are written in terms of the dimensionless fields $a_3$ and $\tilde{a}_8$,  
\be \label{a3a8}
{\bf A}_3({\bf r}) = \frac{a_3(r)}{r}{\bf e}_\varphi  \, , \qquad \tilde{\bf A}_8({\bf r}) = \frac{\tilde{a}_8(r)}{r}{\bf e}_\varphi  \, .
\ee
Inserting Eqs.\ (\ref{phi}) and (\ref{a3a8}) into the potential (\ref{U123}) and separating the contribution of the 
homogeneous CFL phase yields the free energy per unit length of a CFL flux tube, 
\be \label{FL}
\frac{F_{\circlearrowleft}}{L} = \pi \rho_{\rm CFL}^2 \int_0^\infty dR\, R \, U_{\circlearrowleft}(R) \, , 
\ee
where $\rho^2_{\rm CFL}=\mu^2/[\lambda(1-2\eta)]$ is the condensate (squared) in the homogeneous CFL phase, i.e., far away from the flux tube, where we have introduced the dimensionless radial coordinate $R=r\sqrt{\lambda}\rho_{\rm CFL}$, and where the dimensionless free energy density of a flux tube is given by
\bea \label{Uflux}
&&U_{\circlearrowleft}=\frac{\lambda(a_3'^2+\tilde{a}_8'^2)}{R^2} + f_1'^2+f_2'^2+f_3'^2 +\frac{(1-f_1^2)^2}{2}+\frac{(1-f_2^2)^2}{2}+\frac{(1-f_3^2)^2}{2} \nonumber \\[2ex]
&& +f_1^2\frac{N_1^2}{R^2}+f_2^2\frac{N_2^2}{R^2}+f_3^2\frac{N_3^2}{R^2} -\eta\Big[(1-f_1^2)(1-f_2^2)+(1-f_1^2)(1-f_3^2)+(1-f_2^2)(1-f_3^2)\Big] \, . \hspace{1cm}
\eea
Here we have introduced the dimensionless profile functions 
\be
f_i(R) = \frac{\rho_i(R)}{\rho_{\rm CFL}} \, , 
\ee
and have abbreviated
\be \label{N123}
N_1(R) \equiv n_1+\frac{g}{2}a_3(R)+\tilde{g}_8\tilde{a}_8(R) \, , \quad N_2(R) \equiv n_2-\frac{g}{2}a_3(R)+\tilde{g}_8\tilde{a}_8(R)
\, , \quad N_3(R)\equiv n_3-2\tilde{g}_8\tilde{a}_8(R) \, .
\ee
We have not included the trivial $\tilde{\bf B}^2$ contribution into the free energy density (\ref{Uflux}). 
The flux tube profiles are computed from the equations of motion
\bea \label{eom}
a_3''-\frac{a_3'}{R}&=&\frac{g}{2\lambda}\left(f_1^2N_1-f_2^2N_2\right) \, , \\[2ex]
\tilde{a}_8''-\frac{\tilde{a}_8'}{R}&=&\frac{\tilde{g}_8}{\lambda}\left(f_1^2N_1+f_2^2N_2
-2f_3^2N_3\right) \, , \\[2ex]
0&=& \Delta f_1+f_1(1-f_1^2)-f_1\frac{N_1^2}{R^2} -\eta f_1(2-f_2^2-f_3^2) \, , \\[2ex]
0&=& \Delta f_2+f_2(1-f_2^2)-f_2\frac{N_2^2}{R^2} -\eta f_2(2-f_1^2-f_3^2) \, , \\[2ex]
0&=& \Delta f_3+f_3(1-f_3^2)-f_3\frac{N_3^2}{R^2} -\eta f_3(2-f_1^2-f_2^2) \, , 
\eea
where $\Delta$ is the radial part of the Laplace operator in cylindrical coordinates, $\Delta f_i = f_i''+\frac{f_i'}{R}$.
The boundary conditions are $f_i(0)=0$ if $n_i\neq 0$, else $f_i'(0)=0$; $f_i(\infty)=1$, $a_3(0)=\tilde{a}_8(0)=0$, $a_3'(\infty)=\tilde{a}_8'(\infty)=0$. We solve these equations numerically 
with a successive over-relaxation method and insert the result into (\ref{FL}) to obtain the free 
energy of a CFL flux tube. This yields a finite result only if the baryon circulation vanishes. Before we discuss the numerical results without baryon circulation, let us discuss 
the asymptotic behavior of the flux tubes for the general case.

\section{Asymptotic behavior}
\label{sec:asym}

To find the behavior of the profiles far away from the flux tube, we first notice that by evaluating all equations at $R\to \infty$ we find $N_1(\infty) = N_2(\infty)=N_3(\infty)=m/3$ with  
\be
m\equiv n_1+n_2+n_3 \, .
\ee
With the ansatz $a_3(R) = a_3(\infty) + R w_3(R)$, $\tilde{a}_8(R) = \tilde{a}_8(\infty) + R \tilde{w}_8(R)$ for the gauge fields and $f_i(R)=1+u_i(R)$ for the scalar fields we find the 
linearized equations of motion,
\bea
\Delta w_3-\frac{w_3}{R^2}\left(1+\frac{R^2}{\kappa_3^2}\right) &\simeq& \frac{gm}{3\lambda R}(u_1-u_2) \, , \label{v3lin}\\[2ex]
\Delta\tilde{w}_8-\frac{\tilde{w}_8}{R^2}\left(1+\frac{R^2}{\tilde{\kappa}_8^2}\right) &\simeq& \frac{2\tilde{g}_8m}{3\lambda R}(u_1+u_2-2u_3) \, , 
\eea
and 
\bea
\Delta u &\simeq& Mu+\frac{m^2}{9R^2}\left(\begin{array}{c} 1+u_1 \\[2ex] 1+u_2\\[2ex] 1+u_3 \end{array}\right) 
+\frac{m}{3R} \left(\begin{array}{c} gw_3+2\tilde{g}_8\tilde{w}_8\\[2ex] -gw_3+2\tilde{g}_8\tilde{w}_8\\[2ex]
-4\tilde{g}_8\tilde{w}_8\end{array}\right) \, , \label{ulin}
\eea
where, in analogy to a single-component superconductor, we have introduced the Ginzburg-Landau parameters 
for the two gauge fields 
\be \label{kappas}
\kappa_3 \equiv \sqrt{\frac{2\lambda}{g^2}} \, , \qquad \tilde{\kappa}_8 \equiv \sqrt{\frac{\lambda}{6\tilde{g}_8^2}} =\sqrt{\frac{2\lambda}{g^2+\frac{4}{3}e^2}}  \, ,
\ee
and we have defined 
\be
M\equiv 2\left(\begin{array}{ccc}1&-\eta&-\eta\\-\eta&1&-\eta\\-\eta&-\eta&1\end{array}\right) \, , \qquad u \equiv \left(\begin{array}{c} u_1\\u_2\\u_3\end{array}\right) \, .
\ee
Notice that besides $w_3,\tilde{w}_8,u_i\ll 1$ this linearization also assumes $w_3^2,\tilde{w}_8^2 \ll u_i$. 
It is convenient to rotate the scalar fields such that $M$ is diagonalized. We write $\bar{u} = U^{-1}u$ with $U$ such that 
$U^{-1}MU = {\rm diag}(\nu_1, \nu_2, \nu_2)$, with the eigenvalues of $M$
\be
\nu_1\equiv 2(1-2\eta) \, , \qquad \nu_2\equiv 2(1+\eta) \, . 
\ee 
With the weak-coupling results (\ref{weak}), $\nu_1=4$, $\nu_2=1$. 
This transforms 
Eqs.\ (\ref{v3lin}) -- (\ref{ulin}) into 
\bea
\left(\Delta -\frac{1}{\kappa_3^2}-\frac{1}{R^2}\right)w_3 &\simeq& -\frac{gm}{3\lambda R}(\bar{u}_2+2\bar{u}_3) \, \label{eqv3} \\[2ex]
\left(\Delta -\frac{1}{\tilde{\kappa}_8^2}-\frac{1}{R^2}\right)\tilde{w}_8 &\simeq& -\frac{2\tilde{g}_8m}{\lambda R}\bar{u}_2 \, , \label{eqv8}\\[2ex]
\left(\Delta-\nu_1-\frac{m^2}{9R^2}\right)\bar{u}_1 &\simeq& \frac{m^2}{9R^2} \, , \label{equ1}\\[2ex]
\left(\Delta-\nu_2-\frac{m^2}{9R^2}\right)\bar{u}_2 &\simeq& - \frac{4\tilde{g}_8m}{3R}\tilde{w}_8 \, , \label{equ2}\\[2ex]
\left(\Delta-\nu_2-\frac{m^2}{9R^2}\right)\bar{u}_3 &\simeq& -\frac{m}{3R}(gw_3-2\tilde{g}_8\tilde{w}_8) \, . \label{equ3}
\eea
If the three winding numbers add up to zero, $m=0$, i.e., in the case of a pure magnetic flux tube without baryon circulation, all right-hand sides are zero and the equations completely decouple. They have solutions in terms of modified Bessel functions
of the second kind which, allowing for the terms proportional to $m^2$ on the left-hand sides to be nonzero, are
\be
w_3 = c_3 K_1(R/\kappa_3) \, , \quad \tilde{w}_8 = c_8 K_1(R/\tilde{\kappa}_8) \, , \quad \bar{u}_1 = d_1 K_{m/3}(\sqrt{\nu_1}R) \, , \quad \bar{u}_{2,3} = d_{2,3} K_{m/3}(\sqrt{\nu_2}R) \, , \label{bessel}
\ee
with constants $c_3$, $c_8$, $d_1$, $d_2$, $d_3$, 
and the leading asymptotic behavior is obtained from 
\be
K_\alpha(x) = \sqrt{\frac{\pi}{2x}} e^{-x}\left[1+\frac{4\alpha^2-1}{8x}+\frac{(4\alpha^2-1)(4\alpha^2-9)}{2!(8x)^2} + \ldots \right] \, . 
\ee
Undoing the rotation with $U$ then yields the asymptotic behavior of the scalar fields, see first column in Table \ref{tab:n1n2n3}.

\begin{table}[t]
\begin{center}
\begin{tabular}{|c||c|c|} 
\hline
\rule[-1.5ex]{0em}{4ex} 
  & $\;$ zero circulation, $m=0$ $\;$  & nonzero circulation, $m\neq 0$ \\[0ex] \hline\hline
\rule[-1.5ex]{0em}{4ex} 
$w_3$ & $\displaystyle{e^{-R/\kappa_3}}$ & $e^{-R\,{\rm min}(\sqrt{\nu_2},1/\kappa_3)}$ \\[0ex] \hline
\rule[-1.5ex]{0em}{4ex} 
$\tilde{w}_8$ & $\displaystyle{e^{-R/\tilde{\kappa}_8}}$ & $\displaystyle{e^{-R\,{\rm min}(\sqrt{\nu_2},1/\tilde{\kappa}_8)}}$ \\[0ex] \hline
\rule[-1.5ex]{0em}{5ex} 
$u_1,u_2,u_3$ & $\displaystyle{e^{-R\,{\rm min}(\sqrt{\nu_1},\sqrt{\nu_2})}}$ &
$\displaystyle{\frac{m^2}{9\nu_1 R^2}}$ \\[1.2ex] \hline
\rule[-1.5ex]{0em}{5ex} 
$F_{\circlearrowleft}/L$  & finite & $\;$ $\displaystyle{\pi\rho_{\rm CFL}^2\frac{m^2}{3}\ln\frac{R}{R_0}}$ + finite 
$\;$\\[1.2ex] \hline
\end{tabular}
\caption{Leading asymptotic behavior for gauge fields and scalar fields with and without 
baryon circulation, in terms of the dimensionless 
radial coordinate $R$. The gauge fields become trivial, $w_3=0$ and $\tilde{w}_8=0$, if 
$n_1-n_2= 0$ and $n_1+n_2-2n_3= 0$, respectively. The last row shows the logarithmic divergence of the free energy per unit length for the case $m\neq 0$. 
}
\label{tab:n1n2n3}
\end{center}
\end{table}

Let us now discuss the case with baryon circulation, $m\neq 0$. In this case, the equation for 
$\bar{u}_1$ (\ref{equ1}) remains decoupled, and with a power-law ansatz one easily obtains  
\be \label{u1}
\bar{u}_1 = -\frac{m^2}{9\nu_1 R^2}\left[1+\frac{36-m^2}{9\nu_1R^2}+\frac{(36-m^2)(144-m^2)}{81\nu_1^2R^4}+\ldots\right] \, .
\ee
This is the usual behavior of a superfluid vortex, which is expected since $\bar{u}_1$ does not couple to any gauge field. Next, consider Eqs.\ (\ref{eqv8}) and (\ref{equ2}), which couple $\tilde{w}_8$ and $\bar{u}_2$. To determine the leading asymptotic behavior we need to distinguish the cases $\nu_2>1/\tilde{\kappa}_8^2$ and $\nu_2<1/\tilde{\kappa}_8^2$. Let us 
start with the former. In this case, if the right-hand sides of  Eqs.\ (\ref{eqv8}) and (\ref{equ2}) were zero, such that the solutions (\ref{bessel}) would hold, $\bar{u}_2$ would be suppressed stronger than $\tilde{w}_8$. We can retain the asymptotic behavior of the less suppressed function, $\tilde{w}_8\propto R^{-1/2} e^{-R/\tilde{\kappa}_8}$. Inserting this expression 
into the left-hand side of Eq.\ (\ref{eqv8}) yields a result of order $R^{-5/2}e^{-R/\tilde{\kappa}_8}$. We are thus allowed to make an ansatz for $\bar{u}_2$ at this order without violating Eq.\ (\ref{eqv8}), $\bar{u}_2 = d_2 R^{-3/2}e^{-R/\tilde{\kappa}_8}$. This ansatz, being less suppressed than $\bar{u}_2$ would be in the absence of a baryon circulation, is needed to fulfill Eq.\ (\ref{equ2}), from which 
we obtain a relation between $d_2$ and $c_8$. We proceed analogously for the case $\nu_2<1/\tilde{\kappa}_8^2$ to obtain
\bea
\nu_2>1/\tilde{\kappa}_8^2: \qquad \tilde{w}_8 &\simeq& c_8\sqrt{\frac{\pi\tilde{\kappa}_8}{2R}}e^{-R/\tilde{\kappa}_8} \, , \qquad \bar{u}_2 \simeq \frac{4m\tilde{g}_8}{\nu_2-1/\tilde{\kappa}_8^2}\frac{\tilde{w}_8}{3R} \, , \label{v8u21}\\[2ex]
\nu_2<1/\tilde{\kappa}_8^2: \qquad  \tilde{w}_8 &\simeq& \frac{2m\tilde{g}_8}{1/\tilde{\kappa}_8^2-\nu_2}\frac{\bar{u}_2}{\lambda R} \, , \qquad \bar{u}_2 \simeq d_2\sqrt{\frac{\pi}{2\sqrt{\nu_2}R}}e^{-\sqrt{\nu_2}R} \, .\label{v8u22}
\eea
It remains to insert these solutions into Eqs.\ (\ref{eqv3}) and (\ref{equ3}) to determine the asymptotic behavior for $w_3$ and $\bar{u}_3$. We first observe that the case of a vanishing ${\bf A}_3$ field is obtained from setting $w_3=0$ and
$\bar{u}_3=-\bar{u}_2/2$, which implies $u_1=u_2$ for the unrotated fields. This is the case discussed in Ref.\ \cite{Eto:2009kg}. If $w_3$ is nonzero, let us first 
consider the case $\nu_2>1/\kappa_3^2$. Then, after neglecting $\tilde{w}_8$ and $\bar{u}_2$ in Eqs.\ (\ref{eqv3}) and (\ref{equ3}), these equations have exactly the same structure as just discussed for Eqs.\ (\ref{eqv8}) and (\ref{equ2}). Therefore, in analogy to Eqs.\ (\ref{v8u21}) we find
\bea
\nu_2>1/\kappa_3^2: \qquad w_3 &\simeq& c_3\sqrt{\frac{\pi\kappa_3}{2R}}e^{-R/\kappa_3} \, , \qquad \bar{u}_3 \simeq \frac{mg}{\nu_2-1/\kappa_3^2}\frac{w_3}{3R} \, . \label{v3u31}
\eea
We need to verify that it was correct to neglect $\tilde{w}_8$ and $\bar{u}_2$: since $\kappa_3 > \tilde{\kappa}_8$, see Eq.\ (\ref{kappas}), the case $\nu_2>1/\kappa_3^2$ allows for the two cases $\nu_2>1/\tilde{\kappa}_8^2>1/\kappa_3^2$ and $1/\tilde{\kappa}_8^2>\nu_2>1/\kappa_3^2$. In the first case, $\tilde{w}_8, \bar{u}_2 \propto e^{-R/\tilde{\kappa}_8}$, see Eq.\ (\ref{v8u21}), while in the second case $\tilde{w}_8, \bar{u}_2 \propto e^{-\sqrt{\nu_2}R}$, see Eq.\ (\ref{v8u22}). In both cases, these contributions are 
suppressed stronger than $w_3$, $\bar{u}_3$ from Eq.\ (\ref{v3u31}) and it thus was correct to 
neglect $\tilde{w}_8$ and $\bar{u}_2$. Next, we consider the case $\nu_2<1/\kappa_3^2$. This implies  $\nu_2<1/\kappa_3^2<1/\tilde{\kappa}_8^2$, and thus Eq.\ (\ref{v8u22}) shows that $\tilde{w}_8$, $\bar{u}_2$ are of the same leading order as $w_3, \bar{u}_3$, and we find 
\bea 
\nu_2<1/\kappa_3^2: \qquad  w_3 &\simeq& \frac{mg}{1/\kappa_3^2-\nu_2}\frac{\bar{u}_2+2\bar{u}_3}{3\lambda R} \, , \qquad \bar{u}_3 \simeq d_3\sqrt{\frac{\pi}{2\sqrt{\nu_2}R}}e^{-\sqrt{\nu_2}R} \, .\label{v3u32}
\eea
We can now undo the rotation of the scalar fields to compute $u_i$ for the various cases. We find that, since $\bar{u}_1$ contributes to all scalar fields, the leading asymptotic behavior in the case $m\neq 0$ for all $u_i$ is the power law (\ref{u1}). 
This result, together with the behavior of the gauge fields from (\ref{v8u21}) -- (\ref{v3u32}) 
is summarized in the second column of Table \ref{tab:n1n2n3}. In this table, we have also indicated the behavior of the free energy. From Eq.\ (\ref{Uflux}) we see that the terms proportional to $f_i^2N_i^2$ give a divergent contribution if $N_i(\infty)=m/3$ is nonzero. This divergence is logarithmic, expressed in Table \ref{tab:n1n2n3} in terms of an arbitrary scale $R_0$.  
In realistic systems, this divergence is cut off by the boundary of the system or by the presence of other vortices.

\section{Type-I/type-II transition} 
\label{sec:type}

From now on we shall restrict ourselves to pure magnetic flux tubes, $m=0$, in order to discuss the unconventional type-I/type-II behavior of CFL due to 
its multi-component structure. This is most conveniently done by placing CFL into an external magnetic field $H$. Here and in the following $H$ always refers to an ordinary magnetic field (not a color-magnetic field) because eventually we have in mind applications to neutron stars. 
We shall discuss the critical magnetic fields $H_c$, $H_{c1}$, $H_{c2}$. Here, $H_c$ is the critical magnetic field at which the homogeneous CFL phase, with $\tilde{B}_8$ completely expelled, has the same Gibbs free energy as a homogeneous phase that admits penetration of $\tilde{B}_8$, completely or partially. This phase, which supersedes CFL as we go up in $H$, is either  
the completely unpaired phase ("NOR") or the partially paired 2SC phase (which then, in turn, is superseded by the NOR phase at some larger critical magnetic field). 
One can show that as the external field $H$ is increased and 
if only homogeneous phases are considered, there is a transition from CFL to the normal phase for values of the 
strong coupling constant smaller than the critical value  
$g_c^2=8e^2(1+\eta)/[3(1-8\eta)]$, 
and to 2SC for larger values of $g$; see Ref.\ \cite{Haber:2017oqb} for the corresponding phase diagram. In other words, at $g=g_c$ there is a magnetic field $H$ at which all three phases have the same Gibbs free energy. We see that for $1/8<\eta<1/2$ no real value of $g_c$ exist, which means that in this regime there is a direct transition to the normal phase for all $g$. With the weak-coupling result
$\eta=-1/2$, we have $g_c\simeq 0.16$. 

Next, $H_{c2}$ is the upper critical magnetic field of the CFL flux tube phase, assuming a second order transition to either the NOR or the 2SC phase. We are particularly interested in the point $H_c=H_{c2}$ because in an ordinary superconductor this is the point where the behavior changes from type-I to type-II. For small $g$, where we need to compute $H_{c2}$ for the transition between CFL 
and NOR, $H_c=H_{c2}$ turns out to occur at $\sqrt{2\nu_1}=1/\tilde{\kappa}_8$. For the 
transition between CFL and 2SC, relevant for large $g$, we find that $H_c=H_{c2}$ is equivalent to  $\sqrt{\nu_1\nu_2}\sqrt{g^2+e^2/3}/(\sqrt{2}g)=1/\tilde{\kappa}_8$. 

On the other hand, we can compute the point at which the long-range interaction between the flux tubes changes from repulsive to attractive. In an ordinary single-component superconductor, this point is identical to the point $H_c=H_{c2}$. This calculation relies on the CFL flux tubes only and is thus independent of whether the phase at larger $H$ is the NOR phase or the 2SC phase. We find that the sign of the flux tube interaction changes at ${\rm min} (\sqrt{\nu_1},\sqrt{\nu_2})=1/\tilde{\kappa}_8$ for vanishing $B_3$ field and ${\rm min} (\sqrt{\nu_1},\sqrt{\nu_2})=1/\kappa_3$ if $B_3$ is nonzero. Neither of these critical values coincides
with the point $H_c=H_{c2}$.

Finally, we compute $H_{c1}$, the magnetic field at which it becomes favorable to place a single 
flux tube into the system. This critical field is computed numerically from the free energy of the flux tube,
\be \label{Hc1}
H_{c1} = \frac{1}{\tilde{\Phi}_8\sin\theta} \frac{F_{\circlearrowleft}}{L} \, . 
\ee
While for an ordinary superconductor there is a point where $H_c=H_{c1}=H_{c2}$, we find that in CFL no such point exists. In a more complete calculation, taking into account the full flux tube lattice and not just single flux tubes and their long-range interactions, 
first order transitions, say $H_{c1}'$ and $H_{c2}'$, might occur, restoring this point in the form $H_c=H_{c1}'=H_{c2}'$
\cite{Haber:2017kth,preparation}.

\begin{figure} [t]
\begin{center}
\hbox{\includegraphics[width=0.51\textwidth]{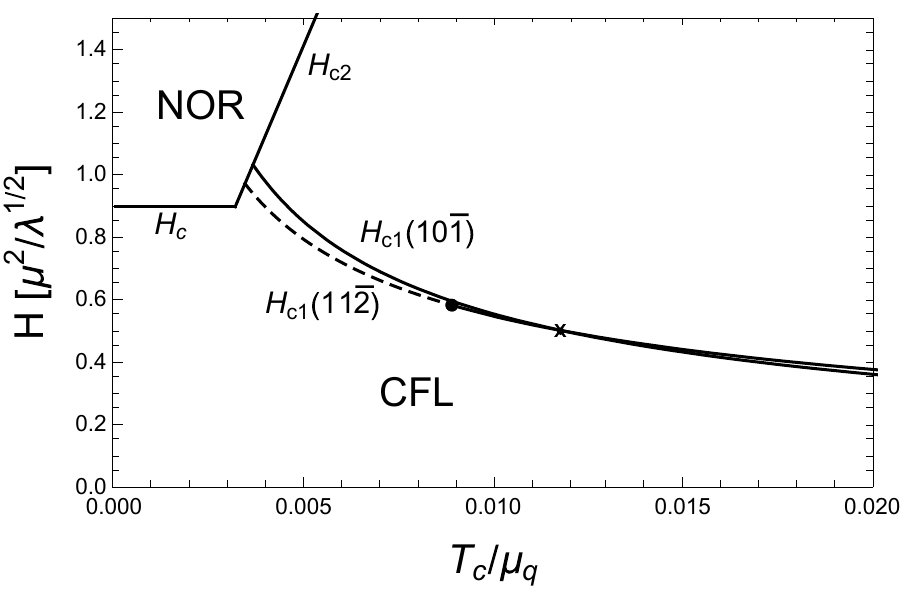}\includegraphics[width=0.49\textwidth]{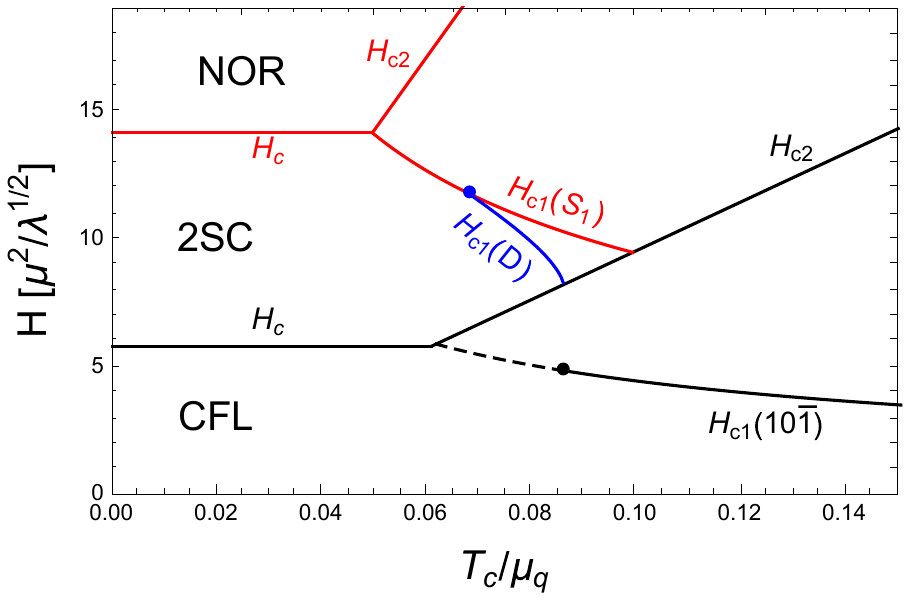}}
\caption{Critical magnetic fields as a function of $T_c/\mu_q$ for $g=0.1$ (left panel) and $g=3.5$ (right panel). In the left panel, the critical magnetic field $H_{c1}$ is shown for the two most preferred windings $(n_1,n_2,n_3)=(1,1,-2)$ and $(1,0,-1)$, the cross indicating the point where these two critical fields intercept. The dashed line is the segment of $H_{c1}$ where the flux tubes attract each other at large distances, such that a first-order transition at a lower magnetic field can be expected to replace the dashed line. The right panel includes the critical fields of the 2SC/NOR
transition, in particular $H_{c1}$ for an ordinary flux tube "$S_1$" and a domain wall "$D$", not discussed in these proceedings. In the right panel, $(n_1,n_2,n_3)=(1,0,-1)$ is the preferred configuration in the entire type-II regime. }
\label{fig:phases}
\end{center}
\end{figure}

In Fig.\ \ref{fig:phases} we show two phase diagrams in the plane of external magnetic field $H$ and $T_c/\mu_q$ for two different values of the strong coupling constant. Both diagrams make use of the weak-coupling parameters (\ref{weak}). From perturbative QCD we know that $T_c/\mu_q$ only depends on $g$ (not explicitly on $\mu_q$) and is exponentially suppressed. Therefore, if $g$ is sufficiently small, the actual value of $T_c/\mu_q$ is basically zero on the linear scale of Fig.\ \ref{fig:phases}, and CFL is a type-I superconductor, independent of the above subtleties.
At large coupling we can no longer trust the perturbative calculation, and due to our lack of knowledge of $T_c$ it makes sense to allow $T_c/\mu_q$ to vary freely, and independently of $g$. 
Extrapolating the weak-coupling result beyond its 
regime of validity one obtains $T_c/\mu_q \simeq 0.01$ for $g=3.5$, which is in the type-I region, but not far from the onset of type-II behavior. Model calculations suggest somewhat larger values
of $T_c$, such that type-II behavior under neutron star 
conditions is conceivable and possibly even the subtleties of the type-I/type-II transition become relevant. 
In Fig.\ \ref{fig:phases} the critical field $H_{c1}$ is shown only for the energetically most 
preferred flux tubes. In the next section 
we present a more systematic calculation, considering all possible combinations of winding numbers (under the constraint $m=0$).

\section{Multi-winding flux tubes} 
\label{sec:multi}

\begin{figure} [t]
\begin{center}
\hbox{\includegraphics[width=0.5\textwidth]{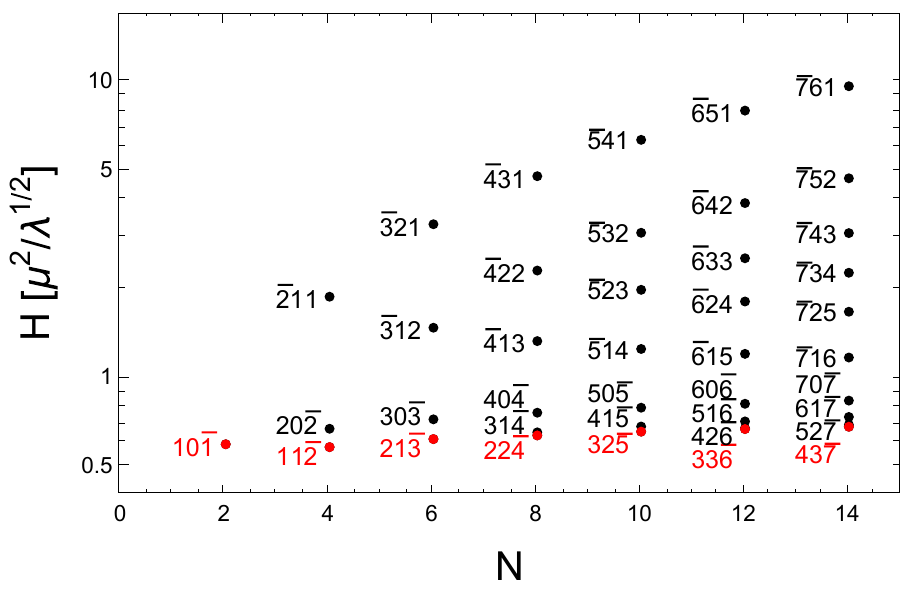}\includegraphics[width=0.5\textwidth]{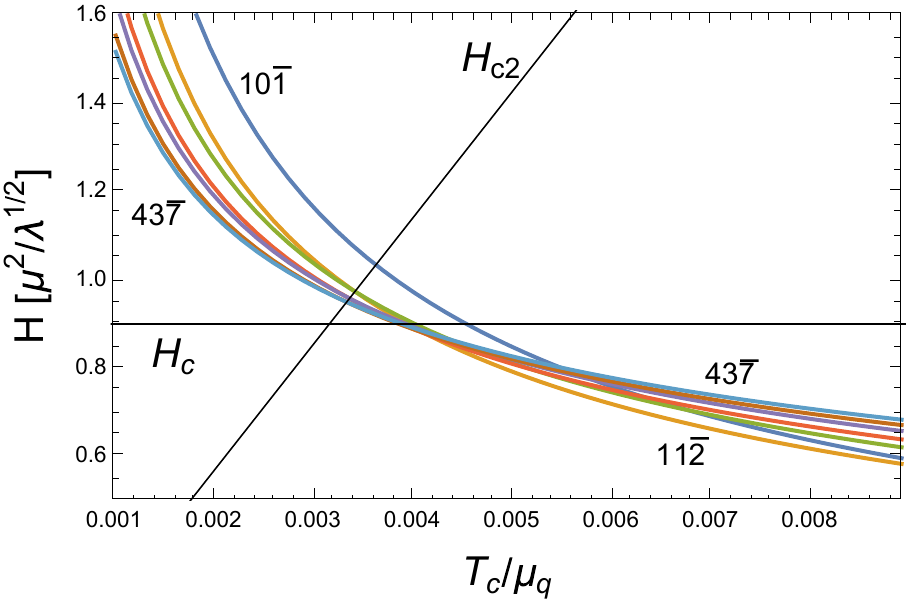}}
\hbox{\includegraphics[width=0.5\textwidth]{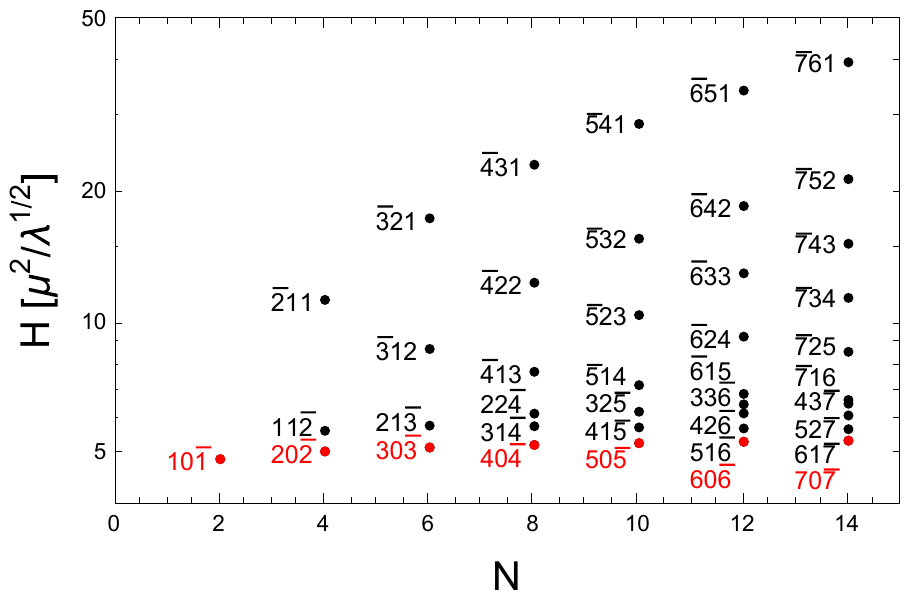}\includegraphics[width=0.5\textwidth]{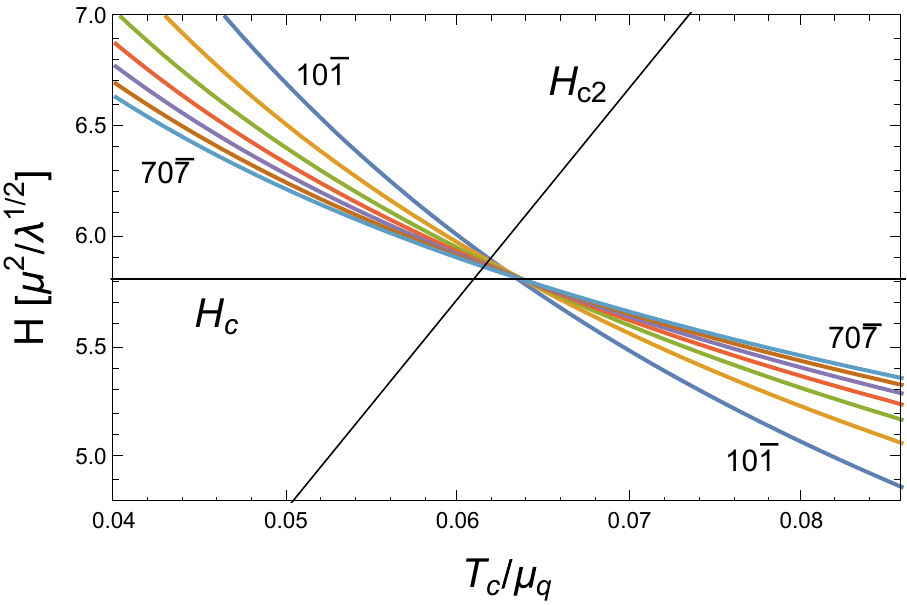}}
\caption{{\it Left panels:} critical field $H_{c1}$ for all flux tube configurations with zero baryon circulation and total winding $N\le 14$ for $g=0.1$ and $T_c/\mu_q\simeq 0.009$ (upper left panel) and $g=3.5$ and $T_c/\mu_q\simeq 0.086$ (lower left panel). The values for $T_c/\mu_q$ correspond to the points where the long-range interaction between flux tubes turns from repulsive to attractive, see Fig.\ \ref{fig:phases}. We indicate negative winding numbers by a bar, for instance $10\bar{1}$ means 
$(n_1,n_2,n_3) = (1,0,-1)$. {\it Right panels:} critical fields $H_{c1}$ for the most preferred configurations for a given $N$ -- marked in red in the left panels -- as a function of $T_c/\mu_q$, for the same values of $g$. The $T_c/\mu_q$ range is chosen such that the value at the right end of the range is the one used in the left panels. We have also plotted the critical fields $H_c$ and $H_{c2}$, indicating transitions to the unpaired phase (upper panel) and the 2SC phase (lower panel); in a single-component superconductor all lines would intersect in the same point.}
\label{fig:Hn}
\end{center}
\end{figure}

In an ordinary, single-component, type-II superconductor, the free energy of a flux tube per magnetic flux is usually lowest for winding number one and increases monotonically with the winding number. In contrast, in a type-I superconductor flux tubes are disfavored, which is reflected in a decreasing 
free energy per magnetic flux with winding number. At the critical point where type-I behavior turns into type-II behavior the free energy per magnetic flux as a function of the winding is flat. In neither regime there is a point at which there is an
energetic minimum at a winding different from one and infinity. This standard situation is already changed by the presence of 
a second component, even if this component is uncharged, as long as it couples to the superconducting component. Nontrivial 
behavior as a function of the winding number in this case was pointed out in Ref.\ \cite{Alford:2007np}. In CFL we not only have three components, but also all three components are charged. Therefore,
a priori we have to deal with three winding numbers. 
To this end, we first define a total winding number by 
\be
N\equiv |n_1|+|n_2|+|n_3| \, .
\ee
Then, we compute the profiles and free energies for all flux tubes with total winding smaller or equal to some total winding $N_0$, which do not carry baryon circulation and which do carry $\tilde{\Phi}_8$ flux. Our 
constraints thus are $m=0$, $n_1+n_2-2n_3\neq 0$, $N\le N_0$. In particular, this collection of 
flux tubes includes configurations with and without $\Phi_3$ flux, i.e., we make no assumption about $n_1-n_2$. In Fig.\ \ref{fig:Hn}, we have chosen $N_0=14$, which leads to 40 configurations, modulo equivalent configurations obtained by
$(n_1,n_2,n_3)\to(-n_1,-n_2,-n_3)$ and $(n_1,n_2,n_3)\to(n_2,n_1,n_3)$. 
The resulting 40 critical magnetic fields $H_{c1}$ are shown in the left panels, for the same two  values of the coupling $g$ as in  Fig.\ \ref{fig:phases}, and two particular choices of $T_c/\mu_q$.  For each of the 7 different $N\le 14$, there is one energetically preferred configuration. In the right panels, $H_{c1}$ as a function of $T_c/\mu_q$ for these 7 configurations is plotted. The main observations from Fig.\ \ref{fig:Hn} are as follows.

\begin{itemize}
\item The gross behavior is as expected: in the type-II regime higher total windings $N$ tend to have larger $H_{c1}$. In particular, in both left panels of the figure we see that $H_{c1}$ increases monotonically with $N$ if for each $N$ we consider the most preferred configuration, except for 
the configuration $11\bar{2}$, which, at weak coupling, is preferred over $10\bar{1}$ in a certain regime of $T_c/\mu_q$, as we already know from Fig.\ \ref{fig:phases}. 

\item For weak coupling, the preferred flux tubes for each total winding are $10\bar{1}$, $11\bar{2}$, $21\bar{3}$, $22\bar{4}$ etc, i.e., they are of the form $xy\bar{k}$ with $k=N/2$ and $x=y$ for 
even $k$ and $x=y+1$ for odd $k$. For strong coupling they are $10\bar{1}$, $20\bar{2}$, $30\bar{3}$, $40\bar{4}$ etc, i.e., they are of the form $k0\bar{k}$. This difference is easy to understand: following the former sequence we create a giant flux tube with 
a completely unpaired core, while the second sequence leads to a giant flux tube with a 2SC core, anticipating the next phase up in $H$ in each case. 

\item As already pointed out in Ref.\ \cite{Haber:2017oqb}, the most relevant configuration for neutron stars appears to be $10\bar{1}$ since it corresponds to the lowest point in the lower left panel. The configuration $11\bar{2}$ \cite{Iida:2004if}, although 
favored for small coupling and although being the only pure flux tube configuration discussed in the literature previous to Ref.\ \cite{Haber:2017oqb}, is, at strong coupling, even disfavored compared to all flux tubes of the form $k0\bar{k}$ shown in the lower left panel.  

\item In a single-component superconductor, the multi-winding curves of the right panels would all intersect in a single point with each other and with $H_c$ and $H_{c2}$. We see that the degree of deviation 
from this standard scenario is different in the two cases shown here. For large coupling, the deviation is small, and the configuration $10\bar{1}$ is always preferred in the 
type-II regime. For weak coupling, the unconventional behavior is particularly obvious, and multi-winding solutions become strong contenders as we approach the type-I/type-II transition region. 

\end{itemize}

\begin{figure} [t]
\begin{center}
\hbox{\includegraphics[width=0.5\textwidth]{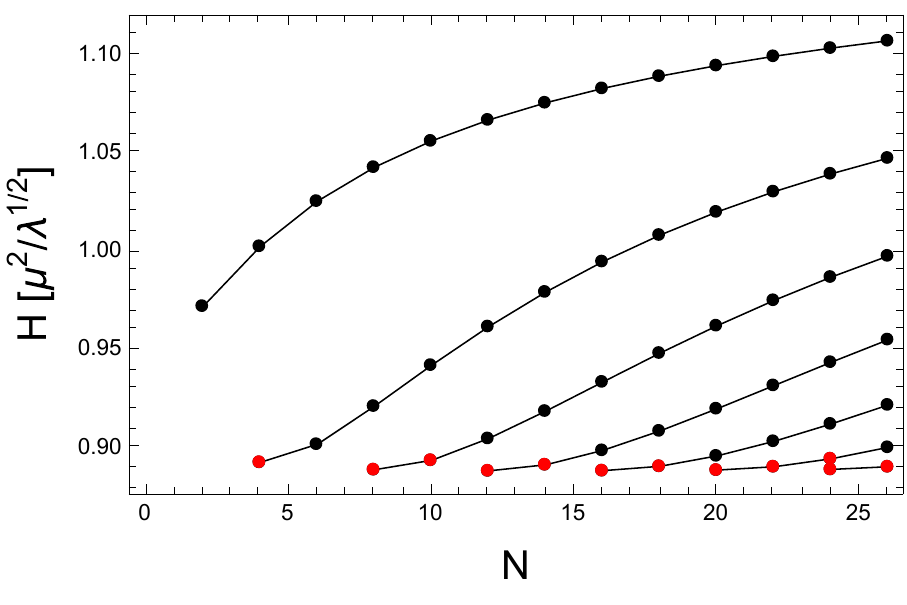}\includegraphics[width=0.5\textwidth]{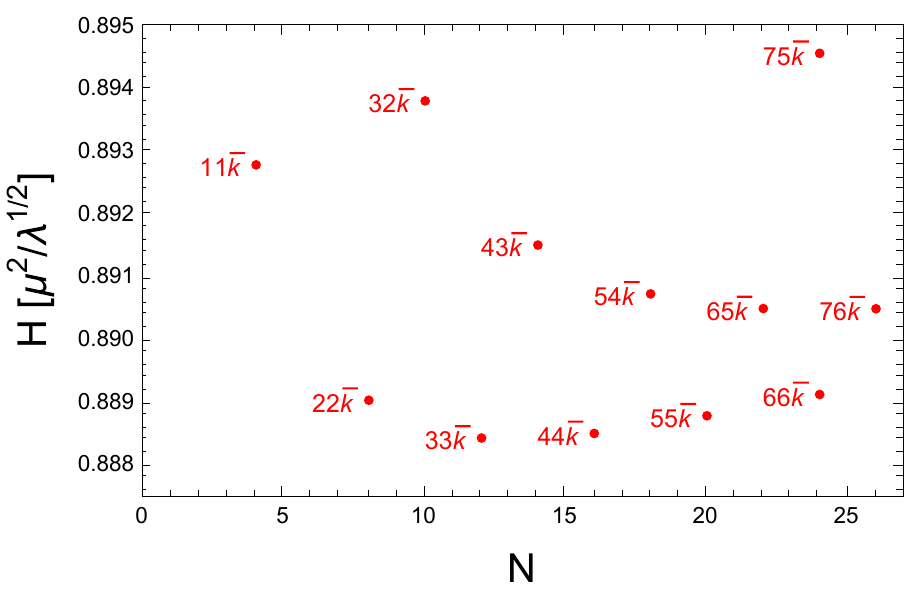}}
\caption{Critical magnetic fields $H_{c1}$ for $g=0.1$ and $T_c/\mu_q=0.004$ as a function of the total winding number $N$. {\it Left panel:} each connected set of points corresponds to one class of winding number triplets, namely $xy\bar{k}$ with $y=0,1,\ldots, 6$ from top to bottom and $k=N/2$, $x=k-y$. {\it Right panel:} Zoom-in to the most preferred configurations, which are marked in red in the left panel. The non-monotonicity with $N$ makes $33\bar{6}$ the favored configuration.}
\label{fig:Hn1}
\end{center}
\end{figure}

To further illuminate the unconventional behavior in the weak-coupling case, we show the critical fields for $g=0.1$ and $T_c/\mu_q=0.004$ in Fig.\ \ref{fig:Hn1}. This figure, in particular the right panel,  requires a good accuracy of the numerics. We have checked that in our relaxation code the points of this panel do not change visibly when we further increase the iteration steps or the number of grid points. 
As a result, we see that for the given value of $T_c/\mu_q$ the multi-winding flux tube $33\bar{6}$ is the preferred configuration. This observation is very interesting, but should be taken with some care for the following reasons. Firstly, in the entire type-II regime shown in Figs.\ \ref{fig:Hn} and \ref{fig:Hn1}, the long-range attractiveness of the flux tubes suggests a first-order phase transition at a lower critical magnetic field than obtained from the single-flux tube free energy (\ref{Hc1}). 
Therefore, the critical fields $H_{c1}$ shown in these figures are not expected to be the actual phase transition lines from a more complete calculation of the phase diagram. Nevertheless, the fact that a multi-winding flux tube has the lowest free energy per magnetic flux in a certain (small) regime suggests that this configuration should be taken into account in such a more complete calculation or in a non-equilibrium situation that allows for metastable states. Secondly, as mentioned above, 
at $g=0.1$, the ratio $T_c/\mu_q$ is most likely much smaller than the value chosen in Fig.\ \ref{fig:Hn1}. Therefore, for QCD this particular result does not appear to be relevant. However, it is not excluded that there is a regime, between the two values $g=0.1$ and $g=3.5$, where the coupling 
is sufficiently small to favor multi-winding solutions and at the same time the critical temperature is sufficiently large to enable type-II behavior. Whether such regime exists is an interesting question that we leave for future studies.  

\begin{acknowledgments}
A.H.\ is partly supported by the U.S. Department of Energy, 
 Office of Science, Office of Nuclear Physics, 
 under Award No.~\#DE-FG02-05ER41375.  A.S.\ is supported by the Science \& Technology Facilities Council (STFC) in the form of an Ernest Rutherford Fellowship. 
 
\end{acknowledgments}

\end{document}